\documentclass[reprint,showpacs,nopreprintnumbers,amsmath,amssymb,superscriptaddress,aps,prl]{revtex4-1} % reprint = approximation to APS final format
\usepackage{color}
\usepackage{graphicx}% Include figure files
\usepackage{bm}% bold math
\usepackage{gensymb} % for generic (text and math mode) scientific symbols like the degree symbol
\usepackage{ifthen}
\begin{document}
\newcounter{wherearethereferences}
\setcounter{wherearethereferences}{5} % 0: references.bib in /home/theo/zwan/articles/references (Kassel home directory), 1: in ../, 2: in current working directory, 3: in ../../../../articles, 4
\hyphenation{acce-le-ra-tion tomo-graphy prio-ri maxi-mum accor-dingly}
\newcommand{\etal}{\textit{et al.\ }}
\newcommand{\mat}[1]{{\mathbi{#1}}}
\newcommand{\vect}[1]{\mathbf{#1}} % for APS: to create bold-faced non-italic Latin and capital Greek symbols
\newcommand{\vectgreek}[1]{\boldsymbol{#1}} % for APS: to create bold-faced (italic) small Greek and special symbols
\newcommand{\Vector}[2]{\genfrac{(}{)}{0pt}{}{#1}{#2}}
\newcommand{\tVector}[2]{\genfrac{(}{)}{0pt}{1}{#1}{#2}} % same as \Vector, but smaller with \textstyle, like \tfrac versus \frac
\newcommand{\iec}{i.e., }
\newcommand{\ie}{\iec}
\newcommand{\egc}{e.g., }
\newcommand{\eg}{\egc}
\newcommand{\egnocomma}{e.g.\ } % American English (for APS)
\newcommand{\di}{d} % \d already defined, for APS
\newcommand{\e}{e} %for APS
\newcommand{\I}{i} %for APS
\newcommand{\eq}{Eq.~} % for APS
\newcommand{\eqs}{Eqs.~} % for APS
\newcommand{\fig}{Fig.~} % for APS
\newcommand{\figs}{Figs.~} % for APS

\newcommand{\thisarticle}{this article~}  

\title{Molecular imaging using high-order harmonic generation and above-threshold ionization}% Force line breaks with \\

\author{Elmar V. \surname{van der Zwan}}
\email{ezwan@itp.uni-hannover.de}
\affiliation{Institut f\"ur Theoretische Physik and Centre for Quantum Engineering and Space-Time Research (QUEST), Leibniz Universit\"at Hannover, Appelstra\ss e 2, D-30167 Hannover, Germany}
\affiliation{Institut f\"ur Physik, Universit\"at Kassel, Heinrich-Plett-Stra\ss e 40, D-31432 Kassel, Germany}
\author{Manfred Lein}
\affiliation{Institut f\"ur Theoretische Physik and Centre for Quantum Engineering and Space-Time Research (QUEST), Leibniz Universit\"at Hannover, Appelstra\ss e 2, D-30167 Hannover, Germany}

\date{\today}% It is always \today, today,

\begin{abstract}
Accurate molecular imaging via high-order harmonic generation relies on
comparing the harmonic emission from a molecule and an
adequate reference system. However, an ideal reference atom with the
same ionization properties as the molecule does not always exist.
We show that for suitably designed, very short laser pulses, a
one-to-one mapping between high-order harmonic frequencies and
electron momenta in above-threshold ionization exists. Comparing
molecular and atomic momentum distributions then provides the
electron return amplitude in the molecule for every harmonic frequency.
We show that the method retrieves the molecular recombination
transition moments highly accurately, even with suboptimal reference atoms.
%High-harmonic generation (HHG) and above-threshold ionization (ATI) are two of the most-studied strong-field processes. So far no practical link between HHG and ATI has been demonstrated. Based on calculations using the strong-field approximation, we show that for extremely short pulses and long laser wavelenghts there exist connections between individual HHG frequencies and ATI momenta. The links between HHG and ATI and their usefulness for molecular imaging are demonstrated using 1D model calculations. From HHG spectra exact molecular matrix element can be retrieved using suboptimal reference atoms if ATI is additionally used to measure the relative ionization rate.
\end{abstract}


\maketitle

When atoms or molecules are irradiated by a strong laser field, high-order harmonic generation (HHG) takes place
and high-frequency photons are emitted \cite{Ferray88}. The interest in HHG from molecules is
growing since observing the radiation is a tool to
investigate the structure of molecules \cite{Lein02,Lein02_2,Itatani04,Haessler10}. 
The sensitivity of the emission spectra to the target structure can be understood
within the three-step model, which provides a semiclassical interpretation of HHG 
in terms of (i) ionization, (ii) free propagation of the electron in the laser field
and return to the parent ion, and (iii) recombination
\cite{Corkum93}. In good approximation, the HHG intensity is proportional
to the modulus squared of the recombination transition dipole moment, or equivalently to the
recombination cross section \cite{Itatani04,vanderZwan08,Lin10}. When the electron continuum
states are additionally approximated as plane waves, one can obtain molecular orbitals
via a tomographic retrieval based on the Fourier transform of the HHG amplitudes
measured from aligned molecules \cite{Itatani04,Haessler10}. These HHG-based molecular imaging
methods rely on the comparison of the harmonic emission to the one from a reference system
with known electronic structure, typically a reference atom. Assuming that the molecule
and the reference atom have the same properties concerning ionization probabilities and
electron propagation, the recombination cross section of the molecule can be isolated.
Clearly, no reference atom with exactly the same ionization properties as the molecule exists. Therefore,
a systematic method to correct for these deviations is highly desirable. 

%For a gas of atoms in a strong laser field, high-harmonic generation (HHG) takes place and high-frequency photons are emitted \cite{Ferray88}. The three-step model provides us with a semi-classical interpretation in terms of ionization, free propagation in the laser field and recombination steps \cite{Corkum93}. HHG can be used as a tool for molecular imaging, for example for determining the internuclear distance \cite{Lein02, Lein02_2} or reconstructing the electronic orbital \cite{Itatani04,Haessler10}. 

Since the probability for recombination of a returning electron is very small, it is likely that 
the system remains ionized, 
and the electron can be detected as an above-threshold ionization (ATI) electron.
ATI momentum distributions have also be used for molecular imaging \cite{Lagmago06,Lin10,Meckel08}. 
It appears plausible to combine HHG and ATI to improve laser-based molecular imaging. 
For reasons outlined in the following, however,
no concrete method has been proposed up to now.

Usually, multi-cycle laser pulses have been used to drive HHG and ATI. This means that
many differerent electron trajectories can potentially contribute to the same harmonic frequency
or to the same electron momentum. In the case of HHG, the same frequency is generated twice per
optical half cycle, namely by the well known short and long trajectory \cite{Lewenstein94}.
In the case of ATI, the interference of contributions from two ionization times
has been termed attosecond double-slit interference \cite{Lindner05}.

%In the usual case of multi-cycle or even semi-infinite (cw) pulses generating HHG and ATI, both HHG and ATI peaks have complex contributions from many different trajectories. 

Initially it was thought there would be a direct correspondence between the HHG and ATI spectra 
(see \cite{Toma99} and references therein) and attempts were made to 
express the harmonic yield as a sum over ATI channels plus recombination 
\cite{Kuchiev99, Kuchiev01}. 
However, no direct link between the intensities of individual HHG and ATI peaks could be drawn 
as in general it is not possible to disentangle the contributions from the different trajectories. 
Two trajectories producing the same harmonic frequency will generally lead to different ATI energies.
Here we show that, for extremely short laser pulses with suitable carrier-envelope phase 
this link turns out to be possible
because only a very limited number of trajectories contributes. 
Taking also advantage of the exponential dependence of the ionization rate on the field strength, 
we present strong one-to-one links from HHG frequencies to ATI momenta,
based on shared birth times of the HHG and ATI trajectories.
We show how to use the relation between HHG and ATI
to improve molecular imaging techniques such as orbital tomography.    

If there is only one trajectory contributing to each ATI momentum and HHG frequency,
and if both trajectories are born at the same time, the ionization steps are identical and
there is a one-to-one mapping from HHG frequency $\omega$ to ATI momentum $p^{(\mathrm{A})}(\omega)$.
We assume here that the ATI electron is emitted along the laser polarization axis, i.e. there
is no rescattering ATI.
Then, the HHG intensity $S(\omega)=\vert \vectgreek{\alpha}(\omega) \vert2$ and 
ATI intensity $A(p)=\vert M(p)\vert2$ are related by
by (atomic units are used throughout) \cite{Corkum93,vanderZwan08,Le09_3}
\begin{subequations}
\label{eq: HHG and ATI yields}
\begin{align}
%\begin{eqnarray}
%S(\omega) &= \vert \vectgreek{\alpha}(\omega) \vert2;
%\quad 
\vectgreek{\alpha}(\omega) &= a(\omega)\; \vect{v}_{\mathrm{rec}}(\omega),
\displaybreak[0]
\\
A(p^{(\mathrm{A})}(\omega)) &= C(\omega)\; \vert a(\omega) \vert2.
\label{eq: schematic A(p)}
\end{align}
\end{subequations}
%\end{eqnarray}
%
Here, the complex amplitude $\vectgreek{\alpha}(\omega)$ is the Fourier transformed dipole acceleration 
and $a(\omega)$ describes the continuum wave packet for HHG. 
The velocity-form recombination matrix element for the HHG process is denoted $\vect{v}_{\mathrm{rec}}(\omega)$. 
The factor $C(\omega)$ relates HHG and ATI and includes the effect of electron
motion after the return time on the momentum distribution. 
Below, we confirm that $C(\omega)$ only depends on the laser field and is independent of the atom or molecule. 
This is in contrast to the quantity $a(\omega)$, which is species-dependent.
Thus, if the momentum distributions $A(p)$ are known for two different systems, the ratio of their
factors $a(\omega)$ can be obtained from Eq. (\ref{eq: schematic A(p)}).

Before demonstrating the improved molecular imaging scheme,
we find suitable laser pulses for which the one-to-one mapping holds. 
%In the ideal case every ATI momentum and HHG frequency is generated by at most one trajectory.
To this end, we express the HHG and ATI yields using the strong-field approximation and 
expand around classical trajectories. For HHG we employ the Lewenstein model \cite{Lewenstein94}.
In this model, the saddle-point integration over momentum gives 
the saddle-point momentum $k_{\mathrm{s}}(t,t')=-\int_{t'}^{t} A(t'') dt''/(t-t')$ 
with $A(t)=-\int_{-\infty}^{t}E(t'')dt''$
such that an electron born at time $t'$ returns to its initial position at recombination time $t$. 
for a linearly polarized laser field $E(t)$.
In contrast to \cite{Chirila08,vanderZwan08}, 
we perform both remaining integrations over $t'$ and $t$ using the saddle-point method. 
The resulting spectrum is generated by trajectories with complex saddle-point times $t'_{\mathrm{s}}$ and $t_{\mathrm{s}}$,
starting with imaginary initial momentum $v_{\mathrm{i}} =\I\sqrt{2 I_{\mathrm{p}}}$ and 
returning with momentum $v_{\mathrm{r}}=\pm \sqrt{2(\omega-I_{\mathrm{p}})}$ where $I_{\mathrm{p}}$ is the ionization potential.
We employ a very short pulse such that all return momenta $v_{\mathrm{r}}$ have the same sign (chosen negative here) 
\cite{vanderZwan07-3}. 
The classical times $t'_0$, $t_0$ are defined by setting $v_{\mathrm{i}} = 0$.
We expand the times $t'_{\mathrm{s}}$, $t_{\mathrm{s}}$ around $t'_0$, $t'_0$ to second order in the Keldysh parameter $\gamma$,
\begin{eqnarray}
\label{eq: expandt}
t_{\mathrm{s}} &= t_0+\frac{1}{2} b_2 \gamma2;\quad t'_{\mathrm{s}} =  t'_0+a_1 \gamma + \frac{1}{2} a_2 \gamma2,\\
a_1 &= \frac{2 \I \sqrt{U_{\mathrm{p}}}}{\vert E(t'_0)\vert };\quad b_2 =\frac{4 U_{\mathrm{p}}}{E(t'_0)(v_{\mathrm{r}}+(t_0-t'_0) E(t_0))},\\
a_2 &= \frac{4 U_{\mathrm{p}}}{(E(t'_0))2}\left(\frac{E(t_0)}{v_{\mathrm{r}}+(t_0-t'_0) E(t_0)}+\frac{E'(t'_0)}{E(t'_0)}\right).
\end{eqnarray}
Expanding to fourth order for the action
$S(t,t') = \frac{1}{2} \int_{t'}^{t} dt'' \: [k_{\mathrm{s}}(t,t')+A(t'')]2 + I_{\mathrm{p}}(t-t')$,
the resulting expression in a \mbox{$d$-dimensional} world is, denoting the bound state as $\psi_0(\vect{r})$
and using $\tau_{\mathrm{s}}=t_{\mathrm{s}}-t'_{\mathrm{s}}$, $\tau_0=t_0-t'_0$,
\begin{eqnarray}
%\begin{subequations}
%\begin{align}
%\begin{split}
\label{eq: SFA2b: alpha(omega)withintegration}
\vectgreek{\alpha}(\omega) =&& -\omega \left(\frac{I_{\mathrm{p}}}{2}\right)^{-\frac{1}{4}} \vect{v}_{\mathrm{rec}}^{*} (v_{\mathrm{r}})  \sum_{t_0,t'_0} \biggl[\frac{2\pi}{\epsilon + \I\tau_{\mathrm{s}}}\biggr]^{\frac{d}{2}}\nonumber\\ 
&&\times d_{\mathrm{ion}}(-\I \:\mathrm{sgn}(E(t'_0))\sqrt{2 I_{\mathrm{p}}},t'_{\mathrm{s}})\nonumber\\ 
&&\times 
 \e^{-\I (S(t_0,t'_0) -\omega t_0)} 
 \e^{
   \I 
%   \left(\frac{E(t_0)}{v_{\mathrm{r}}+\tau_0 E(t_0)}+\frac{E'(t'_0)}{E(t'_0)}\right)
   f(t_0,t'_0)
   \frac{I_{\mathrm{p}}2}{2(E(t'_0))2}
 } \nonumber\\
&&\times \sqrt{
  {\pi }/{\left(
% \left(\frac{E(t_0)}{v_{\mathrm{r}}+\tau_0 E(t_0)}+\frac{E'(t'_0)}{E(t'_0)}\right)
    f(t_0,t'_0)
    \I \sqrt{2 I_{\mathrm{p}}} +\vert E(t'_0)\vert 
  \right)}
}\nonumber\\ 
&&\times \e^{-\frac{(2 I_{\mathrm{p}})^{3/2}}{3\vert E(t'_0)\vert}}
 \sqrt{\frac{2\pi \I\tau_{\mathrm{s}}}{{v_{\mathrm{r}}2}+{\tau_{\mathrm{s}}}E(t_{\mathrm{s}})\:v_{\mathrm{r}}}}+ \mathcal{O}(\gamma3),
%\end{split}
%\end{align}
%\end{subequations}
\end{eqnarray}
with $f(t_0,t'_0)=\frac{E(t_0)}{v_{\mathrm{r}}+\tau_0 E(t_0)}+\frac{E'(t'_0)}{E(t'_0)}$.
The ionization matrix element
\begin{equation}
d_{\mathrm{ion}}(k,t) = \frac{E(t)}{(2\pi)^{d/2}} \int \psi_0(\vect{r}) x\e^{-\I k x} d^d \vect{r},\label{eq: SFA dion}
\end{equation}
exhibits a pole at the saddle-point momentum for Coulombic potentials. 
Supported by the fact that HHG can be modeled succesfully using Gaussian bound states that do not exhibit the pole \cite{Lewenstein94}, 
we replace the integral in \eq\eqref{eq: SFA dion} by an arbitrary constant. 

Similarly, we expand the ATI amplitude $M(p)$ as given by Milo\v{s}evi\'{c} \etal \cite{Milosevic02-2,Milosevic03,Milosevic06} 
around classical birth times. For detailed derivations of \eqs\eqref{eq: expandt}--\eqref{eq: SFA2b: alpha(omega)withintegration} and the 
analogous ATI expression, see \cite{vanderZwanThesisURN}.

\begin{figure}[tbp]	
\center{\includegraphics[width=1.0\columnwidth]{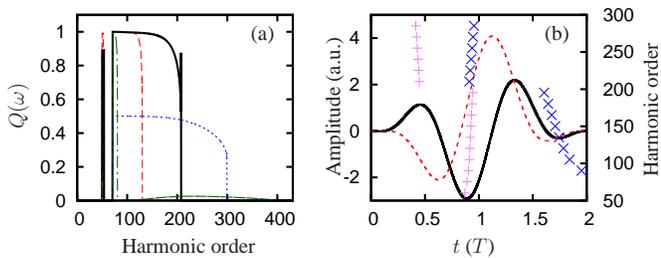}}
\caption{(Color online) (a) $Q(\omega)$ for two-cycle $\sin2$-pulses with $\phi_{\mathrm{CEP}}=\pi$ (red dashed line), $\phi_{\mathrm{CEP}}=1.25\pi$ (black solid line), $\phi_{\mathrm{CEP}}=1.5\pi$ (blue dotted line) and $\phi_{\mathrm{CEP}}=1.75\pi$ (green dot-dashed line); (b) $E(t)$ (x40, black solid line) and $A(t)$ (red dashed line) for $\phi_{\mathrm{CEP}}=1.25\pi$. Also indicated are the birth (violet plusses) and recombination (blue crosses) times of the dominant trajectories.}
\label{fig: Q1to1h}
\end{figure}

We calculate the uniqueness of a trajectory in determining harmonic $\omega$ by dividing the absolute value of the corresponding term in \eq\eqref{eq: SFA2b: alpha(omega)withintegration} by the total sum.
Similarly, for every harmonic trajectory we also calculate the uniqueness of the ATI trajectory born at the same time in determining its associated ATI momentum. The maximum attainable product of these two factors is labeled $Q(\omega)$. 
For the maximum possible value $Q=1$, there is a perfect correspondence between a harmonic frequency and ATI momentum through their shared birth time. 
%If $Q=0$, no link between HHG and ATI is possible at all. 
In \fig\ref{fig: Q1to1h}(a) we show $Q(\omega)$ for different two-cycle $\sin2$-laser pulses with intensity $I=2\times 10^{14}$ W/cm$^2$ and wavelength $\lambda=2000$ nm shining on a 1D system with the ionization potential set to $I_{\mathrm {p}}=30.2$ eV. The pulses are characterized by the carrier-envelope phase $\phi_{\mathrm{CEP}}$, \ie the phase between the envelope and the carrier wave of the pulse. 
We only consider ATI momenta whose amplitudes are greater than $1\,$a.u. 

A two-cycle pulse with $\phi_{\mathrm{CEP}}=1.25\pi$ gives rise to a good link between HHG and ATI over a broad harmonic range. This result is nearly independent of the dimensionality $d$ (not shown).
In \fig\ref{fig: Q1to1h}(b) we plot the time-dependent electric field and vector potential of this pulse, 
and we indicate the birth and recombination times of the dominant trajectories. 
Because $\vert E(t) \vert$ is much higher during the birth time of harmonic orders $\sim 100\text{--}200$ ($t\simeq 0.92T$) than around $t\simeq 1.35T$---the only other time where ATI electrons with the same final momentum are born---the link between HHG and ATI arises. 
Smartly selecting experimental 
phase-matching conditions might allow somewhat longer pulses to provide useful links between HHG and ATI.
However, for slightly longer pulses the HHG frequencies are linked to very low ATI momenta which 
may require including Coulomb corrections for the classical trajectories.
%are strongly influenced by the Coulomb potential. 
%Including Coulomb corrections for the classical trajectories and/or 

We verify the link between HHG and ATI by numerical solution of the time-dependent Schr\"odinger equation (TDSE) for 1D H$_2^+$ 
with varying internuclear distance $R$. We use the softcore potential 
\begin{equation}
V(x) = \frac{-Z}{\sqrt{(x-\frac{R}{2})2+a2}} - \frac{Z}{\sqrt{(x+\frac{R}{2})2+a2}},
\end{equation} 
where the softness parameter $a2$ is adjusted such that $I_{\mathrm{p}}=30.2$ eV. 
The TDSE is solved on a grid using the split-operator method \cite{Fleck76,Feit82}, and the bound states are found by imaginary-time 
propagation \cite{Kosloff86}. The grid length is 24027 a.u.\ and it contains 143360 grid points. After the end of the laser pulse, 
the wave function is propagated for two more cycles. 
The HHG spectrum is calculated from a windowed Fourier transform of the dipole acceleration and 
the ATI spectrum is obtained from the momentum-space representation of the wave function 
after removing the bound states by windowing out the inner 40 a.u. in position space.
\begin{figure}[tbp]	
\center{\includegraphics[width=1.0\columnwidth]{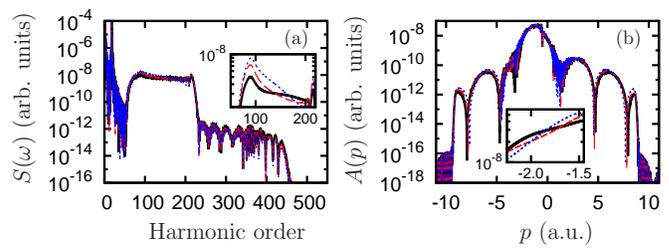}}
\caption{(Color online) (a) HHG spectra  for 1D H$_2^+$ at $R=2.00$ a.u.\ (black solid lines), $R=2.03$ a.u.\ (red dashed lines) and $R=2.06$ a.u.\ (blue dotted lines). (b) ATI momentum distributions. The insets show smoothed details of the spectra.}
\label{fig: spectra}
\end{figure} 
For the ground state with $Z=0.731$, the HHG spectra and ATI momentum distributions 
for three different internuclear distances are shown in \fig\ref{fig: spectra}. 
Here we employ the pulse of \fig\ref{fig: Q1to1h}(b).
%with $\phi_{\mathrm{CEP}}=1.25\pi$ at an intensity of $2\times 10^{14}$ W/cm$^2$ and a wavelength of 2000 nm.   

\begin{figure}[tbp]	
\center{\includegraphics[width=0.8\columnwidth]{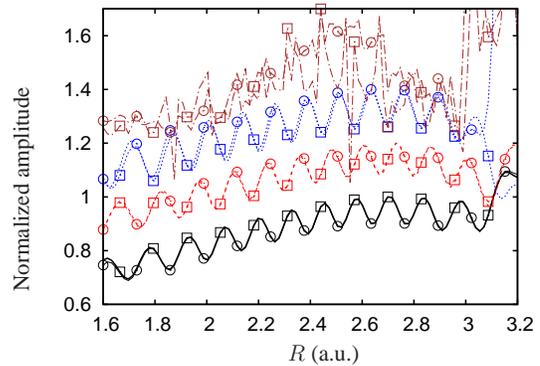}}
\caption{(Color online) Vertically aligned $\vert a(\omega) \vert$ (lines with circles) and $\sqrt{A(p^{(\mathrm{A})}(\omega))}$ (lines with squares) as a function of $R$. We consider harmonic orders 100 (black solid lines), 150 (red dashed lines), and 200 (blue dotted lines) and their associated ATI momenta $p^{(\mathrm{A})}= -1.4922$ a.u., $p^{(\mathrm{A})}=-1.8301$ a.u., and $p^{(\mathrm{A})}=-2.2315$ a.u., respectively. The curves were shifted vertically for clarity. Also shown are harmonic 150 and ATI momentum $p^{(\mathrm{A})}=-1.8301$ a.u.\ from a six-cycle trapezoidal pulse with 1-cycle ramps (brown dot-dashed lines), for comparison. }
\label{fig: HHGATIlink}
\end{figure}

Similar to the quantitative rescattering theory \cite{Le09_3,Lin10}, we calculate the 1D recombination matrix elements $v_{\mathrm{rec}} = \langle \psi_0(x) \vert x \vert \psi_{\mathrm{s}}(x) \rangle$ using field-free scattering states $\psi_{\mathrm{s}}$. We obtain numerically exact $\psi_{\mathrm{s}}$ by integrating the static Schr\"odinger equation using the Numerov method (see \egnocomma \cite{Blatt67}) on a grid with a total length of 4000 a.u.\ and 320000 grid points. For an electron approaching from positive $x$ we set the wave function equal to $\exp(-\I k(\omega) x)$ for the two lowest grid points, where $k(\omega)=\sqrt{2(\omega-I_{\mathrm{p}})}$. After integrating upwards, we require that for large positive $x$ the wave function is given by 
\begin{equation}
c\,\psi_{\mathrm{s}}(x) = e^{-\I k(\omega) x} + R_{\mathrm{e}} e^{\I k(\omega) x},
\end{equation}  
where $R_{\mathrm{e}}$ is the reflection coefficient. This leads to a normalization constant 
\begin{equation}
c=2 e^{-\I k(\omega) x}/\left(\psi_{\mathrm{s}}(x)+\I \psi_{\mathrm{s}}'(x)/k\right).
\end{equation} 
In \fig\ref{fig: HHGATIlink} we demonstrate the link between HHG and ATI as a function of $R$ using the other parameters of \fig\ref{fig: spectra}. 
For three different harmonic orders we plot $\vert a(\omega) \vert$ and $\sqrt{A(p^{(\mathrm{A})}(\omega))}$ (both normalized) using the links between HHG and ATI obtained from \eqs\eqref{eq: SFA2b: alpha(omega)withintegration} and the corresponding ATI expression. The remarkable overlap between HHG and ATI that only breaks down for the largest $R$ confirms the strong link between HHG and ATI indicated by the black solid line in \fig\ref{fig: Q1to1h}(a). The ratios of normalization constants correspond to $\sqrt{C(\omega)}$ (see \eq\eqref{eq: schematic A(p)}). The oscillation in $R$ of both the HHG and ATI amplitudes in \fig\ref{fig: HHGATIlink} is caused by excitation to the first excited state before ionization. Between two maxima the first excited state drops in energy by exactly $\omega$. We have verified that the oscillation period is inversely proportional to the laser wavelength prior to the moment of ionization. Both the fact that the HHG and ATI curves for a six-cycle pulse are less similar to each other and the fact that they exhibit wild behavior are related to multiple trajectories contributing to the yields. 

\begin{figure*}[tbp]	
\center{\includegraphics[width=0.8\textwidth]{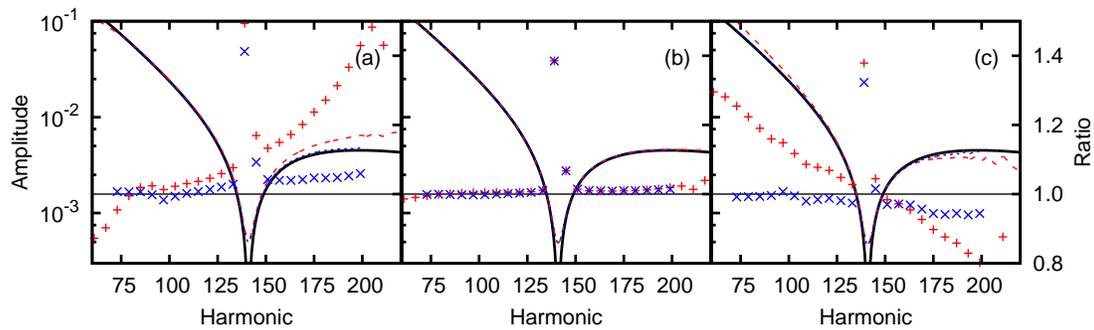}}
\caption{(Color online) Exact matrix element $v_{\mathrm{rec}}(\omega)$ for the first excited state of H$_2^+$ (black solid lines), recovered matrix element using only HHG (red dashed lines) and recovered matrix elements using HHG and ATI (blue dotted lines). Also shown is the ratio of the recovered to the exact matrix elements on a linear scale using only HHG (red plusses) and using HHG and ATI (blue crosses); (a) $Z^{(a)}=1.4$ a.u., (b) $Z^{(a)}=2.6$ a.u., (c) $Z^{(a)}=4$ a.u. }
\label{fig: dexact retrieval}
\end{figure*}

The link between HHG and ATI gives the experimentalist access to the ratio of the instanteneous ionization rates of different molecules during the high-harmonic generation process, and as such is a useful tool in studying HHG and molecular imaging. In particular, the estimate for the continuum wave packet needed for the tomographic reconstruction of molecular orbitals \cite{Itatani04} can be improved using
\begin{equation}
a_\theta(\omega) = a^{(a)}(\omega)\sqrt{A_\theta(p^{(\mathrm{A})}(\omega)) / A^{(a)}(p^{(\mathrm{A})}(\omega))}.
\label{eq: a(k) with ATI)}
\end{equation} 
Here $\theta$ is the orientation of the molecule in the laser field and with the superscript `$(a)$' we denote quantities belonging to the reference atom in the tomographic procedure. Demonstrating numerically the possibility of combined HHG-ATI molecular imaging, we retrieve the field-free matrix elements of the first excited state of 1D H$_2^+$ using a reference atom. We employ
\begin{equation}
v_{\mathrm{rec}}(\omega) \simeq \alpha(\omega)/a(\omega)
\end{equation}
with either $a(\omega)= a^{(a)}(\omega)P_{\mathrm{I}}/P_{\mathrm{I}}^{(a)}$ (HHG imaging), where $P_{\mathrm{I}}$ is the total ionization probability, or with \eq\eqref{eq: a(k) with ATI)} (HHG-ATI imaging). Here we use for 1D H$_2^+$ the parameters $R=2$ and $Z=1.3$ a.u.\ and the same laser pulse as for \fig\ref{fig: spectra}. As the reference atom we use 1D softcore models with different nuclear charges $Z^{(a)}$. For all systems the softcore parameter $a2$ is adjusted such that $I_{\mathrm{p}}=30.2$ eV. The results of the simulation can be found in \fig\ref{fig: dexact retrieval}.
The figure shows that when the total nuclear charge is identical for the molecule and the reference atom (\fig\ref{fig: dexact retrieval}(b)), the molecular matrix element can be accurately retrieved using only HHG from the molecule and atom \cite{Zeidler06}, thereby also demonstrating the accurateness of our field-free matrix elements \cite{Lin10}. However, when the total nuclear charge does not match (\figs\ref{fig: dexact retrieval}(a) and \ref{fig: dexact retrieval}(c)) the propagation step becomes different for the atom and molecule and errors arise in the retrieval of the matrix elements. These errors largely disappear by incorporating also ATI electrons in the retrieval procedure, demonstrating the potential of \eq\eqref{eq: a(k) with ATI)} for orbital tomography and molecular imaging in general. The shallowing of the retrieved matrix elements comes from diffusing the HHG and ATI spectra with Gaussians with $1/e$-widths of $\Delta \omega=6\omega_{\mathrm{L}}$ a.u.\ and $\Delta p =2\sqrt{0.2\omega_{\mathrm{L}}}$ a.u., respectively.

In summary, we used the saddle-point approximation and expansions in $\gamma$ to evaluate strong-field expressions for HHG and ATI in terms of sums over classical trajectories. Using these expressions we have shown that for extremely short laser pulses and long laser wavelengths there exists strong links between individual frequencies and momenta of HHG and ATI. We demonstrated these links and their potential for molecular imaging using 1D model calculations. Future molecular imaging experiments will benefit from this effect. 

\begin{acknowledgments}
The authors thank Ciprian C. Chiril\u{a} for discussions on the saddle-point expression for HHG and the Deutsche Forschungsgemeinschaft for funding the {\em Centre for Quantum Engineering and Space-Time Research} (QUEST). We acknowledge the support from the European Marie Curie Initial Training Network Grant No.\ CA-ITN-214962-FASTQUAST.
\end{acknowledgments} 

\ifthenelse{\arabic{wherearethereferences}=0}{
\bibliography{/home/theo/zwan/articles/references}
}
{
\ifthenelse{\arabic{wherearethereferences}=1}{
\bibliography{../references}
}
{
\ifthenelse{\arabic{wherearethereferences}=2}{
\bibliography{./references}
}
{
\ifthenelse{\arabic{wherearethereferences}=5}{
\bibliography{../../../../../articles/references}
}
{
\ifthenelse{\arabic{wherearethereferences}=4}{
\bibliography{/home/itp/ezwan/scratchdirect/home/articles/references}
}
{
\bibliography{../../../../articles/references}
}
}
}
}
}

\end{document}